\documentclass[a4paper]{spie}  
\usepackage[]{graphicx}
\usepackage[]{amsmath}
\usepackage[]{subfigure}
\title{Modeling charge transport in Swept Charge Devices for X-ray
spectroscopy}

\author{P. S. Athiray\supit{a}\supit{b}, S. Narendranath\supit{a}, P.
Sreekumar\supit{a}, J. Gow\supit{c}, V. Radhakrishna\supit{a}, B.R.S.
Babu\supit{b}\supit{d}
\skiplinehalf
\supit{a}ISRO Satellite Centre, Vimanapura, Bangalore, 560017 India \\
\supit{b}Department of Physics, University of Calicut, Kerala, India\\
\supit{c}Planetary and Space Sciences Research Institute, The Open University, UK\\
\supit{d}Department of Physics, Sultan Qaboos University, Muscat
}

\authorinfo{Further author information: (Send correspondence to
P. S. Athiray)\\P. S. Athiray: E-mail: athray@gmail.com, Telephone: +91 9901818170}

\begin{document}
  \maketitle


\begin{abstract}

We present the formulation of an analytical model which simulates
charge transport in Swept Charge Devices (SCDs) to understand the
nature of the spectral redistribution function (SRF). We attempt to
construct the energy-dependent and position dependent SRF
by modeling the photon
	interaction, charge cloud generation and various loss mechanisms viz.,
	recombination, partial charge collection and split events.
	The model will help in optimizing event selection, maximize event
	recovery and improve
	spectral modeling for
	Chandrayaan-2 (slated for launch in 2014). A proto-type physical model is developed and the
	algorithm along with its results are discussed in this paper.
\end{abstract}




\section{INTRODUCTION}
\label{sec:intro}  

X-rays emitted from the Sun incident on the lunar surface interact with major rock forming
elements, producing X-ray Fluorescence
(XRF) emission. X-ray line energies and intensities of each are used to
study the surface chemistry of the celestial object.
Though similar technique was deployed in earlier missions starting from
Apollo, the Chandrayaan-1 mission was optimally designed to generate the
maximum coverage and generate the most spectroscopically accurate measure
of major elemental abundance. The unusual low solar activity hampered
completion of the primary scientific objective of
Chandrayaan-1 X-ray Spectrometer (C1XS)\cite{Grande09} in creating a global lunar
elemental map using XRF emission lines. 
Chandrayaan-2 Large Area Soft X-ray Spectrometer\cite{rkrish11} (CLASS) is being
developed for the upcoming mission Chandrayaan-2 (slated for launch in 2014) to complete the
objective of global mapping of lunar surface chemistry. To achieve enhanced
sensitivity to detect XRF emission lines from the Moon even during
weak and quiescent solar conditions,  CLASS is designed to have a total
geometric area of 64 $cm^2$ using Swept Charge Device (CCD-236)\cite{Holland08}
developed by e2V Technologies Ltd., UK. It uses an array of 16 SCDs each
with an area of 4 $cm^2$.\\

Details about SCDs are described in Sec. \ref{sec:SCD} along with the architecture of
CCD-54 which is modeled in the current work. Sec. \ref{sec:model} briefly
covers the details of the charge transport in X-ray detectors along with
assumptions used in our model. Understanding and
interpretation of simulation results along with comparison with CCD-54 calibration data
are discussed in Sec. \ref{sec:Response}. Summary of ongoing and future
works are given in Sec. \ref{sec:future}.

\section{Swept Charge Device (SCD)}
\label{sec:SCD}  

SCD's are modified version of conventional
two-dimensional X-ray CCD's. They can be
considered as one-dimensional pseudo linear sensors developed exclusively
for non-imaging, spectroscopic
purposes. Through Multi-Pinned Phase mode operation and rapid continuous
clocking, they provide high spectral resolution within benign
operating temperature range of -20$\,^{\circ}\mathrm{}$C to +5$\,^{\circ}\mathrm{}$C
for X-rays in the 0.5-10 keV range. The
advantages of SCD over a conventional two-dimensional X-ray CCD are many: the
reduction in read-out complexity, large collection area, good spectral
resolution, minimal cooling requirement, suppression of surface generated dark
current and avoidance of image integration period.

\subsection{SCD - CCD-54 - Architecture}
\label{sec:SCDarchitecture}  

\begin{figure}
   \begin{center}
   \begin{tabular}{c}
   \includegraphics[height=8cm]{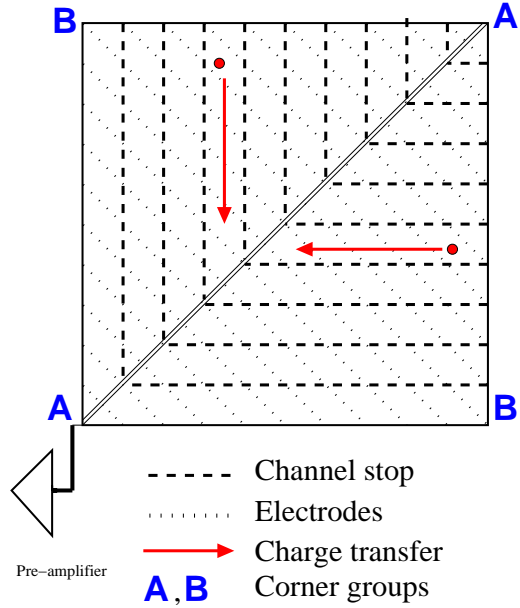}
   \end{tabular}
   \end{center}
   \caption[example]
   { \label{fig:ccd54}
Schematic of CCD-54. The channel stops (dashed lines) and diagonal
electrodes (dotted lines) are depicted respectively. Charge cloud produced by
an X-ray photon will be clocked between the channel stops represented by a
long arrow. Corners of SCD are grouped as A, B whose locations
in the SCD are also marked.}
   \end{figure}

CCD-54 consists of 1725 diagonal electrodes covering an
active area  of 1.07cm$^2$ including inbuilt buried channels.
Schematic view of CCD-54 is shown in Fig. \ref{fig:ccd54}. Channel stops are
fabricated in the epitaxial silicon wafer in the form of a herringbone structure.
Charges produced in the underlying buried channel are clocked towards the
diagonal transfer channel and then towards the readout amplifier.
CCD-54 is
configured for 3 ${\phi}$ clocking and hence need 575 transfers to flush the
complete device. A detailed description about the device is given in Lowe et al.\cite{Lowe01}

\subsection{Spectral Redistribution Function (SRF) of SCD's}
\label{sec:SRF}  
Absorption of an X-ray photon in a detector results in a complex cascade
of energy transfers. Energy deposited by the photon gets transformed in many
ways leading to various signatures in the observed energy spectrum. The
distribution of energy deposits observed in a device for an incident
mono-energetic photon is called the spectral redistribution function (SRF).
SRF strongly depends on the energy of the incident
photon. It is observed that the SRF of SCD's exhibit four distinct features\cite{Narendranath10}
viz., a low energy tail, a low energy shelf, a low peak and escape peak in
addition to the photopeak. Some of these feature
arise due to the coupled action of diagonal readout and photon interaction
at different depths in the SCD. Hence it is useful to have a detailed
physical model to understand the SRF of SCD's from which contributions by
each component can be derived. Development of a physical model of the SRF will help in:

\begin{enumerate}
	\item {Optimizing the logic for selection of events in SCD.}
	\item {Recovering non-photopeak events which leads to a gain in the
		overall efficiency of the detector system.}
\end{enumerate}
Typically, SRF of pixellated detectors may be pixel position dependent.
SRF of corner pixels differ significantly
from the other pixels due to charge losses. We also study the pixel
location dependency of SRF separately. For this purpose, the four corner pixels are grouped
into two based on the similarity of SRFs. Group A contains two corner pixels which are (i) the one next to
the readout amplifier and its diagonal counter part, marked as A in
Fig. \ref{fig:ccd54}. The other two corner pixels form Group B (marked as B in
Fig. \ref{fig:ccd54}).

\section{Charge transport model}
\label{sec:model}  
The fundamental difference between a traditional X-ray CCD and SCD lie in its
diagonal clocking readout structure\cite{Jason09}. In CCD's, the amount of
charge collected in each pixel due to photon interaction is preserved
during the readout except for small charge transfer losses. In contrast,
the diagonal charge transfer architecture of SCD allows charges  from
different areas of the device to get summed up at the diagonal prior to the final readout. Hence we first modeled a conventional 2-D X-ray CCD in our simulation over which we then incorporated the diagonal clocking readout mechanism of SCD.\\
	\begin{figure}
   \begin{center}
   \begin{tabular}{c}
   \includegraphics[height=8cm]{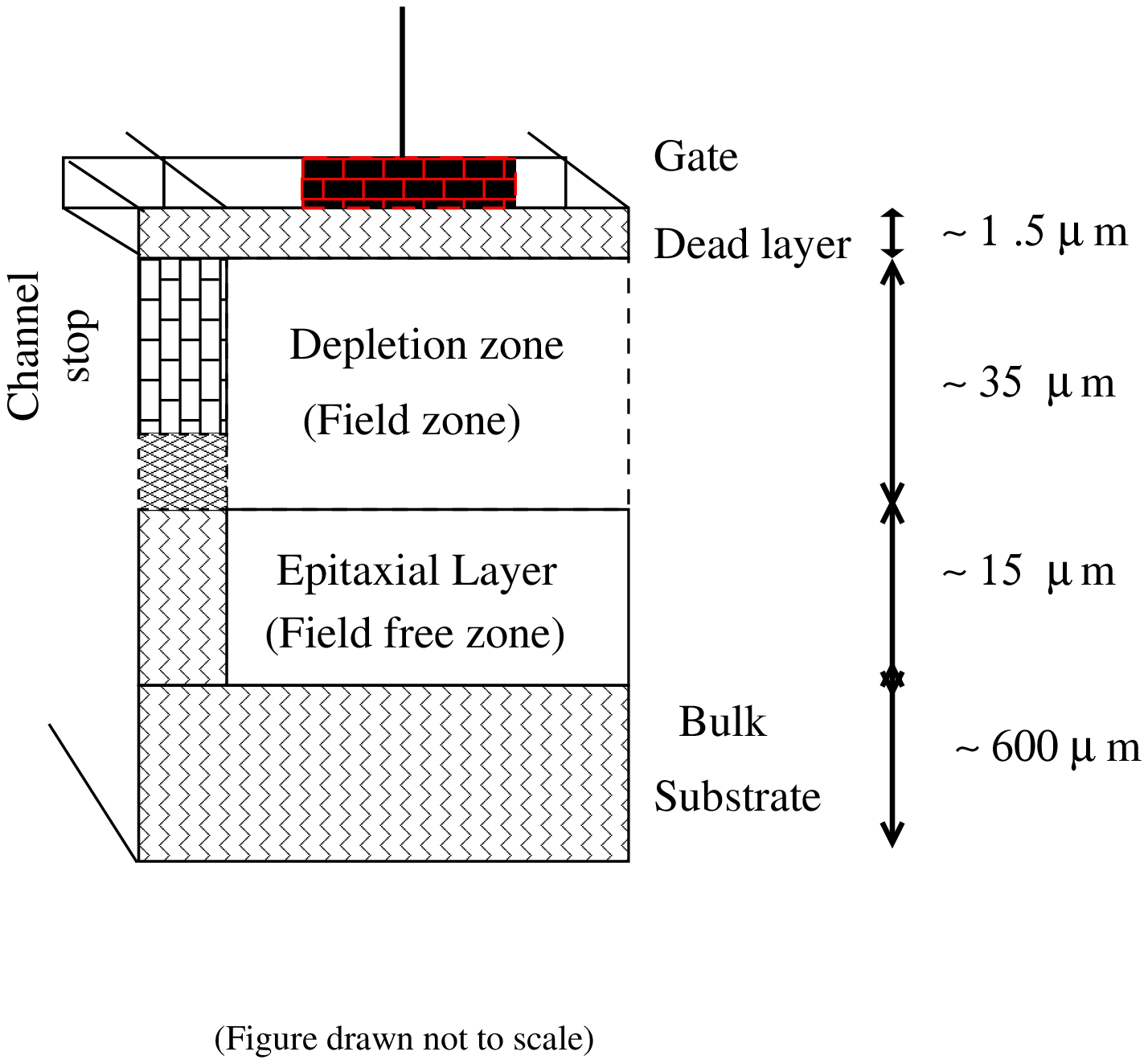}
   \end{tabular}
   \end{center}
   \caption[example]
   { \label{fig:ccd54structure}
The generic structure of CCD-54 device where photons are incident from the top.
Interactions in the non-shaded layers are considered in the current model
which contributes the most to the observed energy response. Approximate
dimensions of different layers used in the code are taken from Narendranath
et al.\cite{Narendranath10}}
   \end{figure}

\subsection{Considerations in the model}
The observed spectral response of SCD arises mainly from
interactions at five different zones viz., channel stop, dead layer,
field zone, field-free zone and bulk substrate. Channel stop occupies
only a very tiny fractional area
of the entire SCD ($<$ few \%). They are considered to be responsible for partial charge collection producing a low energy shelf.
Total thickness of dead layer made up of multiple slabs (SiO$_2$, Poly Si, Si$_3$N$_4$) in SCD is ${\approx}$ 1 to 2 ${\mu}$m.
Dead layer interaction contributing to the observed SRF becomes significant
for low energy X-rays (${\leq}$ 3 keV).  X-ray photons with energies between
3-10 keV exhibit a maximum probability to interact in the field zone,
field-free zone and bulk substrate. Predominant recombination and diffusion
in the bulk zone due to very high doping concentrations ($N_a {\approx}10^{18}
cm^{-3}$) will not allow the charge cloud to reach the collection gate.
Hence in this work, we have modeled only
photon interactions in the field and field-free zones. Photon
interactions in the channel stop and dead layer are expected to be small.\\

\begin{figure}
   \begin{center}
   \begin{tabular}{c}
   \includegraphics[width=10cm]{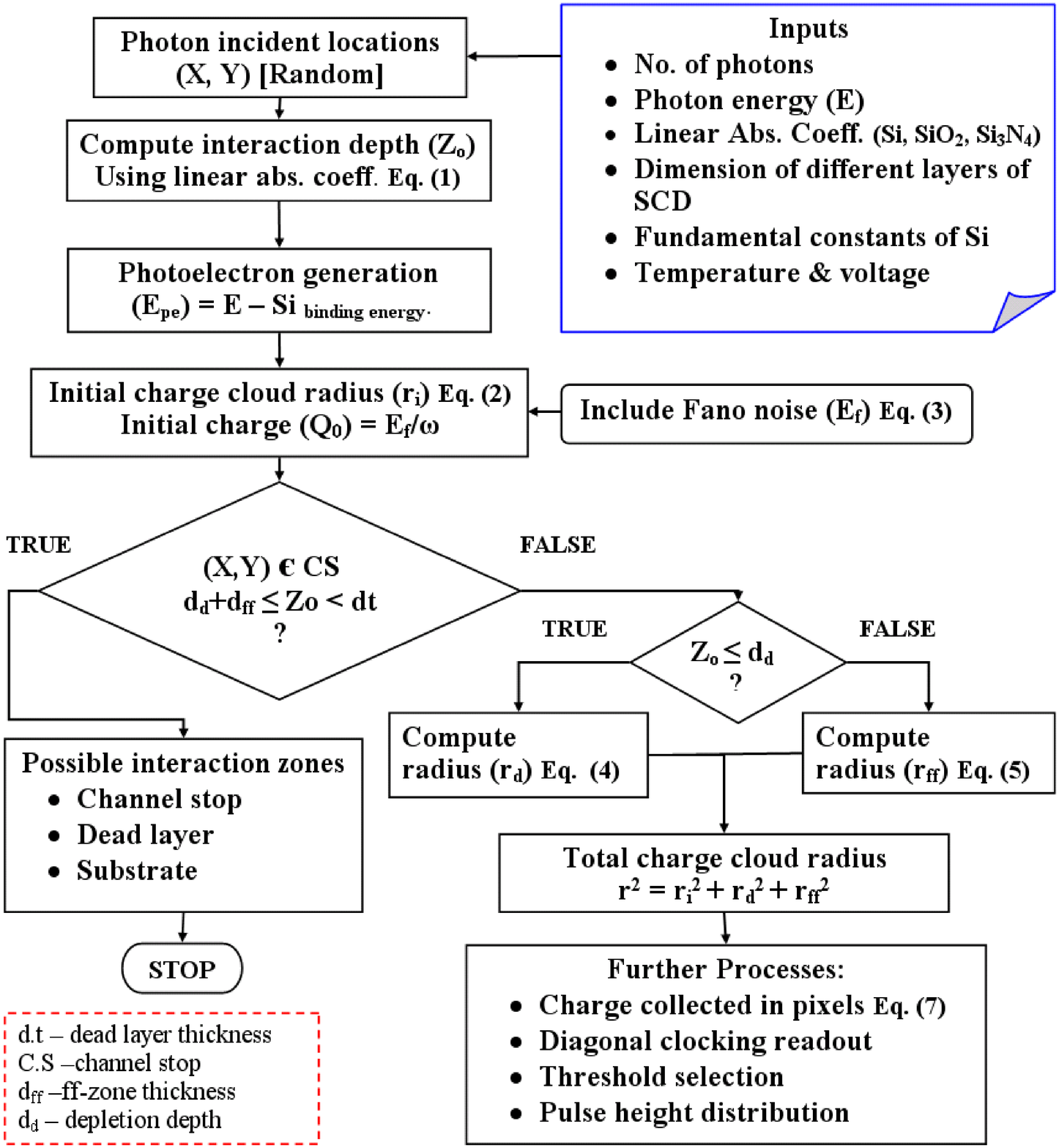}
   \end{tabular}
   \end{center}
   \caption[example]
   { \label{fig:algorithm}
Flowchart explaining the architecture of charge transport model}
     \end{figure}

We followed the approach adopted in simulating the response of Advanced
Chandra Imaging Spectrometer (ACIS)\cite{Townsley02} and X-ray telescope (XRT) in Swift mission
\cite{Godet09}. We follow a Monte-Carlo based algorithm to
simulate the interaction of mono-energetic X-ray photons in different
layers of SCD. The aim of the model is to understand and quantify the
non-photopeak events seen in the observed spectra. We attempt to
model the photon interaction at different regions of CCD-54 both along the
thickness and across the surface of the device. An algorithm is developed and code is
written in Interactive Data Language (IDL). Global structure of SCD considered in the present model with
different component slabs arranged in order is shown in Fig.
\ref{fig:ccd54structure}. A flowchart explaining the algorithm of current
model is shown in Fig. \ref{fig:algorithm}. \\

\subsection{Photon interactions}
\label{sec:interaction}  
Photons are incident on the SCD from the top and travel through different
layers before it interacts. Depending on the energy of incident X-ray
photons the depth distribution at which interactions occur in the detector is computed by :

	\begin{equation}
	\label{eq:interactiondepth}
	z_o=-\frac{1}{{\mu}(E)} ln(R_u)
	\end{equation}
where ${\mu}(E)$ is the linear mass absorption coefficient of the material at
 photon energy E, $R_u$ is a uniform random number. 
 If the interaction depth is greater than the thickness of a slab, then the
 interaction depth is re-computed for the following slab.
 Absorption of an X-ray photon via photo-electric effect produces a
photo-electron which on further ionization produces a charge cloud
(e$^-$-h pairs). Initial radius ($r_i$) of this assumed spherical charge
cloud is related to the energy of the photo-electron ($E_{pe}$)\cite{Kurniawan07} :
	\begin{equation}
	\label{eq:initialchargecloudradius}
	r_i =\left\{ \begin{array}{ll}
				40.0 \frac{E_{pe}^ {1.75}}{{\rho}} & (5~keV < E_{pe} < 25~keV)\\
							 				& \\
				30.9 \frac{E_{pe}^ {1.53}}{{\rho}} & (E_{pe} {\leq} 5~keV)
			\end{array}
			\right.
	\end{equation}

where ${\rho}$ is the density of the detector material. It is also assumed that the radial charge profile of the initial charge cloud
follows Gaussian distribution. In order to estimate the distribution of charge contained in the cloud, Fano noise is added to the photon's energy using a normal random number generator which is given as :
\begin{equation}
	\label{eq:fanonoise}
	E_f = E_i + R_n(0)\sqrt{F{\omega}E_i}
	\end{equation}
where $E_f$ is the Energy distribution with Fano noise added, $R_n(0)$ is normal distributed random
number with mean 0, F is the Fano factor and ${\omega}$ is the average
energy required to produce an e$^-$-h pair.

\subsubsection{Field \& Field-free zone interactions}
\label{sec:f_ff_int}  
Interactions at depths ($z_0$) within the depletion zone are termed as
field zone interactions. The negative charge cloud produced within the
depletion depth ($d_d$) will experience the electric field and hence will
drift towards the collection anode. Considering linear regime i.e., drift velocity
${\propto}$ electric field, radius of the charge cloud ($r_d$) at the gate
 is computed after drifting through the depletion region.\\

Epitaxial region above the bulk substrate and below the depletion zone is
called field-free zone.
Due to the absence of electric field in this region, the charge cloud suffer
from diffusion and recombination before it reaches the gate. Diffusion in
field-free zone enlarges the
radius of the charge cloud and causes spill over of charges across pixels.
When charges are not contained within a pixel due to a photon hit, the
events are called multi-pixel events or split events which are predominant
for field-free zone interactions. It was clearly demonstrated by Pavlov et
al\cite{Pavlov99}. that the charge density distribution is non-Gaussian for
field-free zone interactions.\\

For simplicity, we assume the charge cloud
follows a Gaussian distribution and compute the
radius of the charge cloud ($r_{ff}$) reaching the interface between field
and field-free zone. Once the charge cloud reaches the boundary
of field zone it drifts in the electric field and reaches the
collection gate. Final radius ($r$) of the charge cloud at the gate is computed
by taking the quadrature sum of $r_i$, $r_d$, $r_{ff}$ from which the
amount of charge collected in each pixel is derived. Standard fundamental equations
used in the model for the computation of radius of charge clouds and charge
collection in pixels are given in Appendix A. Values of some of the
important parameters used in the model are listed in Table~\ref{tab:parameters}.

\begin{table}[h]
\caption{Values of parameters used in modeling charge transport of CCD-54}
\label{tab:parameters}
\begin{center}
\begin{tabular}{|l|l|} 
\hline
\rule[-1ex]{0pt}{3.5ex} Voltage (for $d_d$ computation) & 3.8 V  \\
\hline
\rule[-1ex]{0pt}{3.5ex}  Temperature & 263 K  \\
\hline
\rule[-1ex]{0pt}{3.5ex}  Channel stop pitch & 25 ${\mu}$m  \\
\hline
\rule[-1ex]{0pt}{3.5ex}  Number of acceptor impurities ($N_a$) & 4 ${\times}
10^{12} cm^{-3}$  \\
\hline
\rule[-1ex]{0pt}{3.5ex}  Number of pixels & 25 ${\times}$ 25  \\
\hline
\rule[-1ex]{0pt}{3.5ex}  Number of mono-energetic photons & 3${\times}10^{4}$  \\
\hline

\end{tabular}
\end{center}
\end{table}

\section{Response Simulation}
\label{sec:Response}  
Using the algorithm described in Sec. \ref{sec:model}, we simulated the SRF of SCD for multiple scenarios of photon interactions. In
this section, we present the simulation results along with our
understanding of the derived SRF. Simulations are performed for
large number of different mono-energetic photons (4 keV, 4.51 keV, 5.0 keV, 6.0 keV and 8.0
keV). Simulations include interactions in (i)
field zone and (ii) field-free zone at different pixel locations. As the laboratory calibration
data are available at 4.51 keV (K$_{\alpha}$ of Ti), one of the simulation is performed
at 4.51 keV.\\
   \begin{figure}
   \begin{center}
   \begin{tabular}{c}
   \label{figur}
   \subfigure{\label{fig:f_cent}\includegraphics[height=6cm]{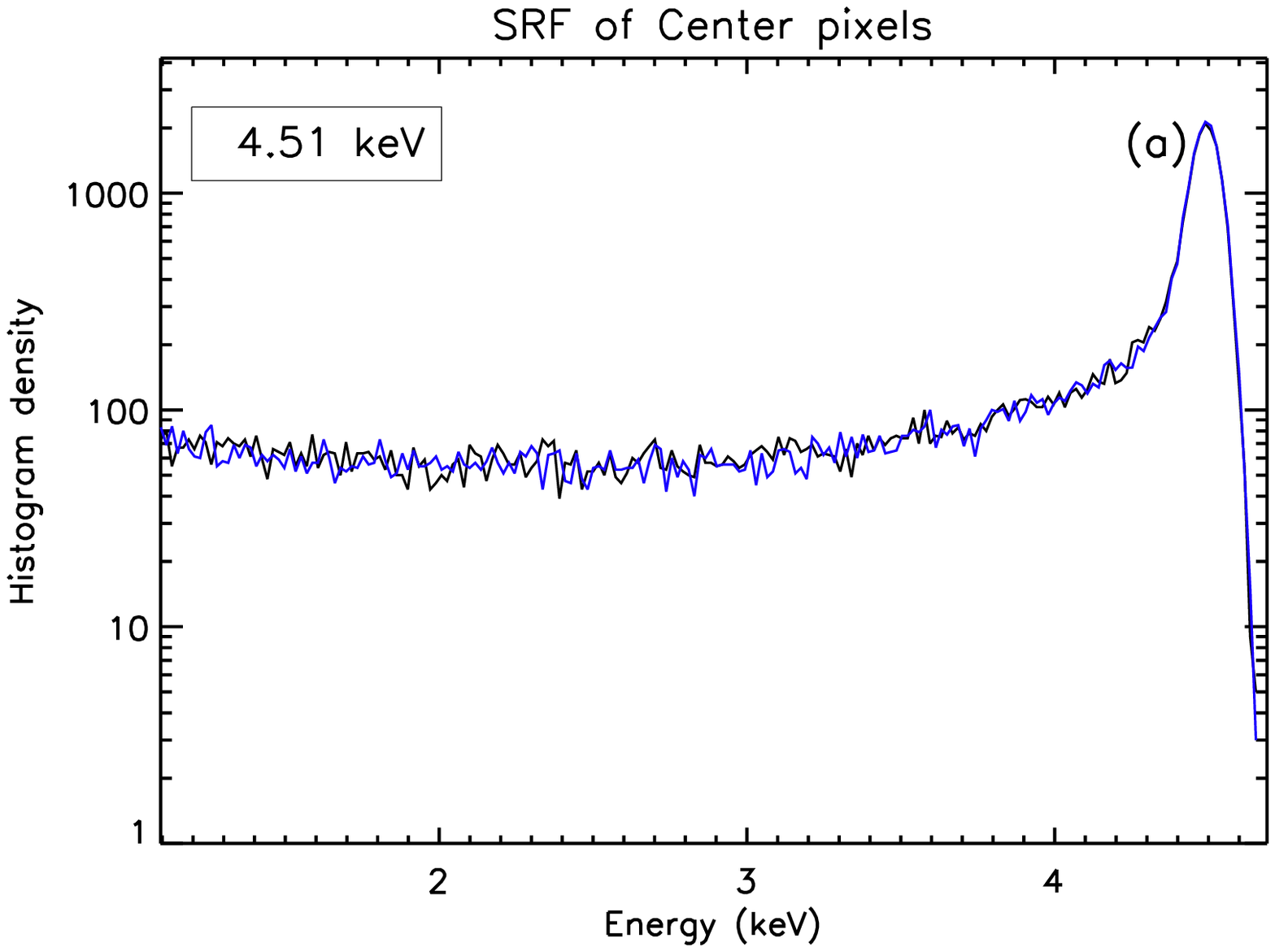}}
   \subfigure{\label{fig:f_corn}\includegraphics[height=6cm]{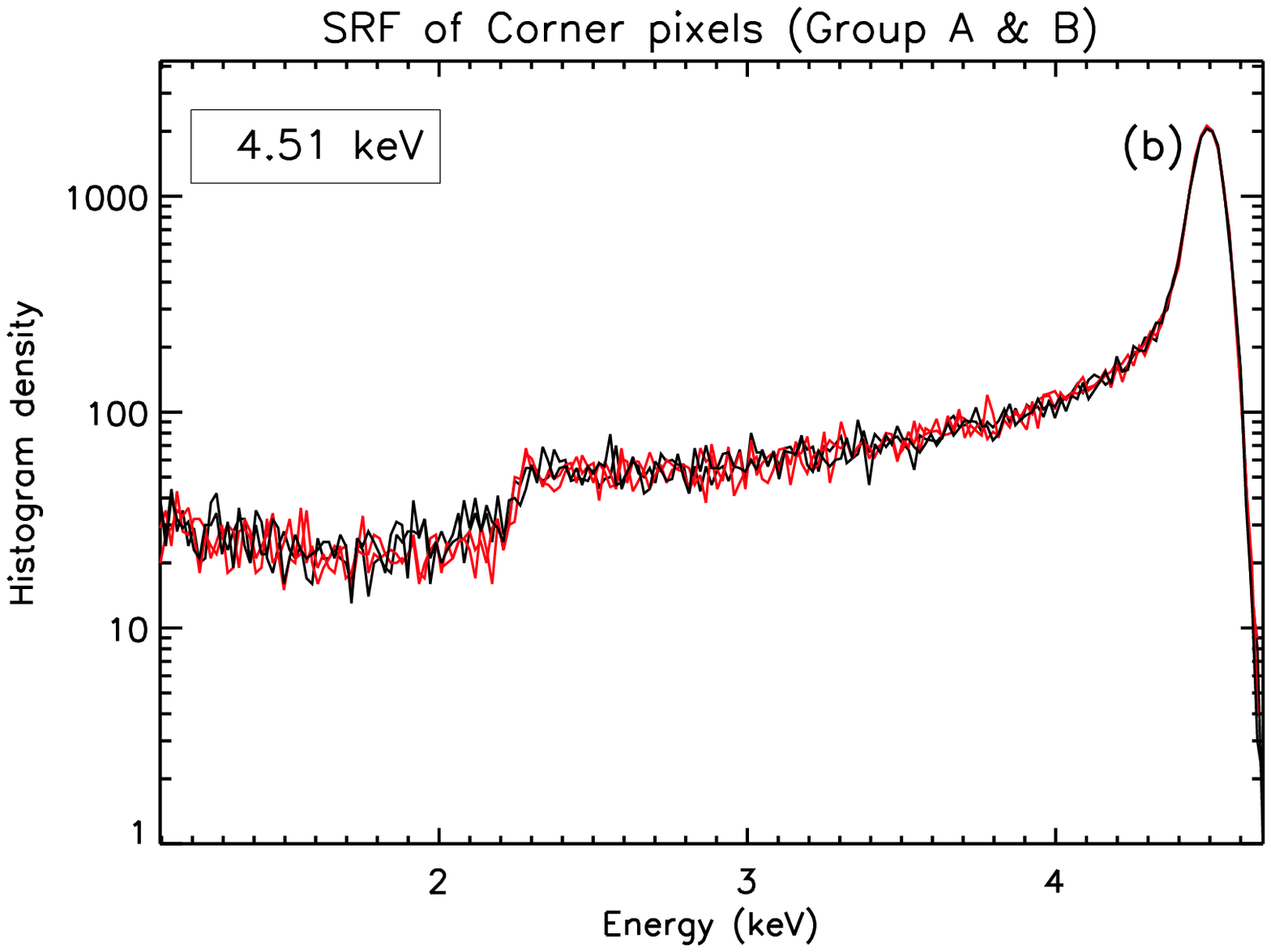}}
   \end{tabular}
   \end{center}
   \caption[example]
   { \label{fig:srffieldzone}
Simulated SRF of CCD-54 for 4.51 keV photons interacting in the field
zone (a) photon incidence on center pixels (b) photon incidence on corner
pixels (both group A (black) \& B (red)). Photopeak and non-photopeak (low energy tail
component) are clearly seen. The difference in the SRF is due to charge loss
in the corner pixels which produces the dip seen in the low energy tail component.}
   \end{figure}

  \begin{figure}
   \begin{center}
   \begin{tabular}{c}
   \subfigure{\label{fig:ff_cent}\includegraphics[height=4cm]{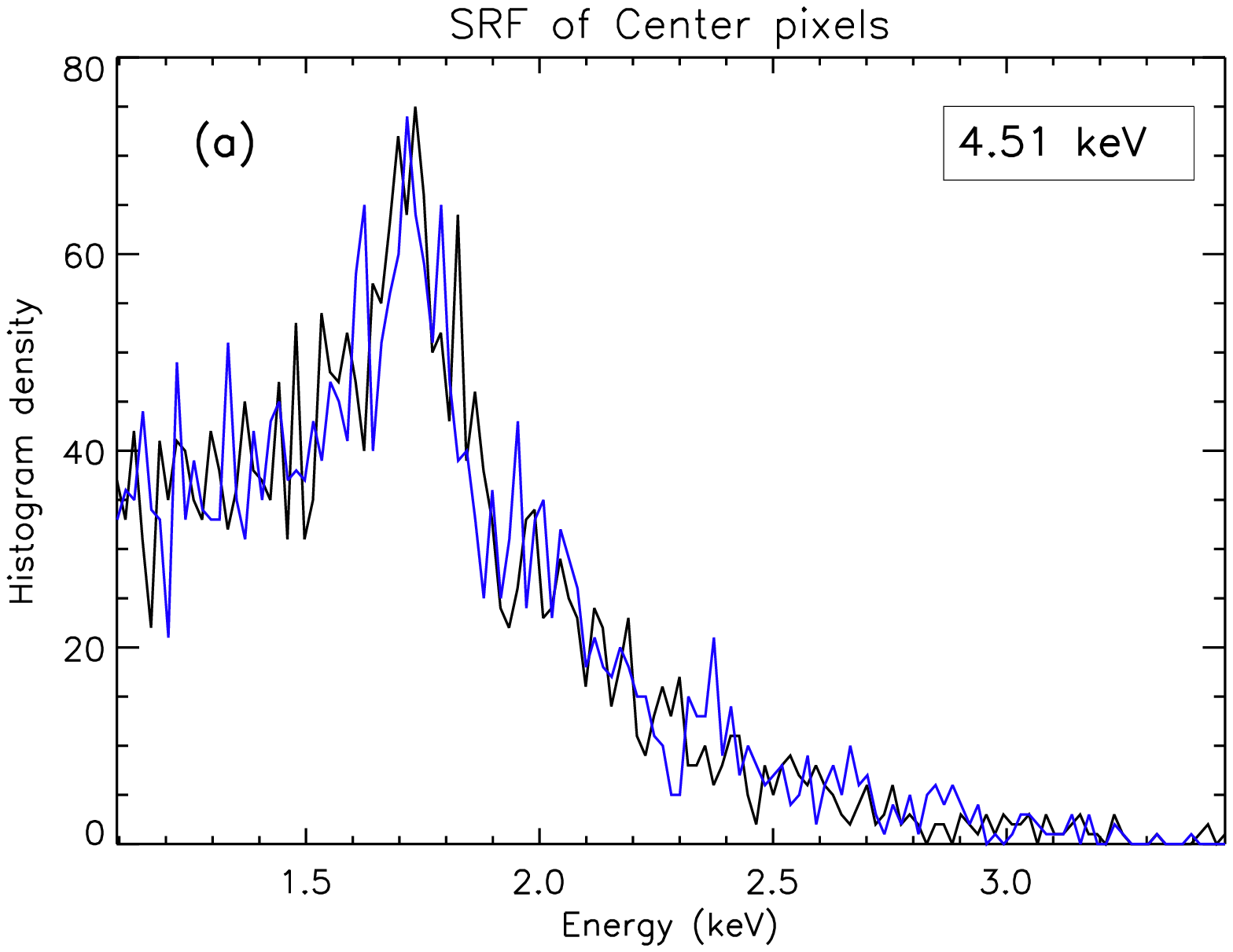}}
   \subfigure{\label{fig:ff_cornA}\includegraphics[height=4cm]{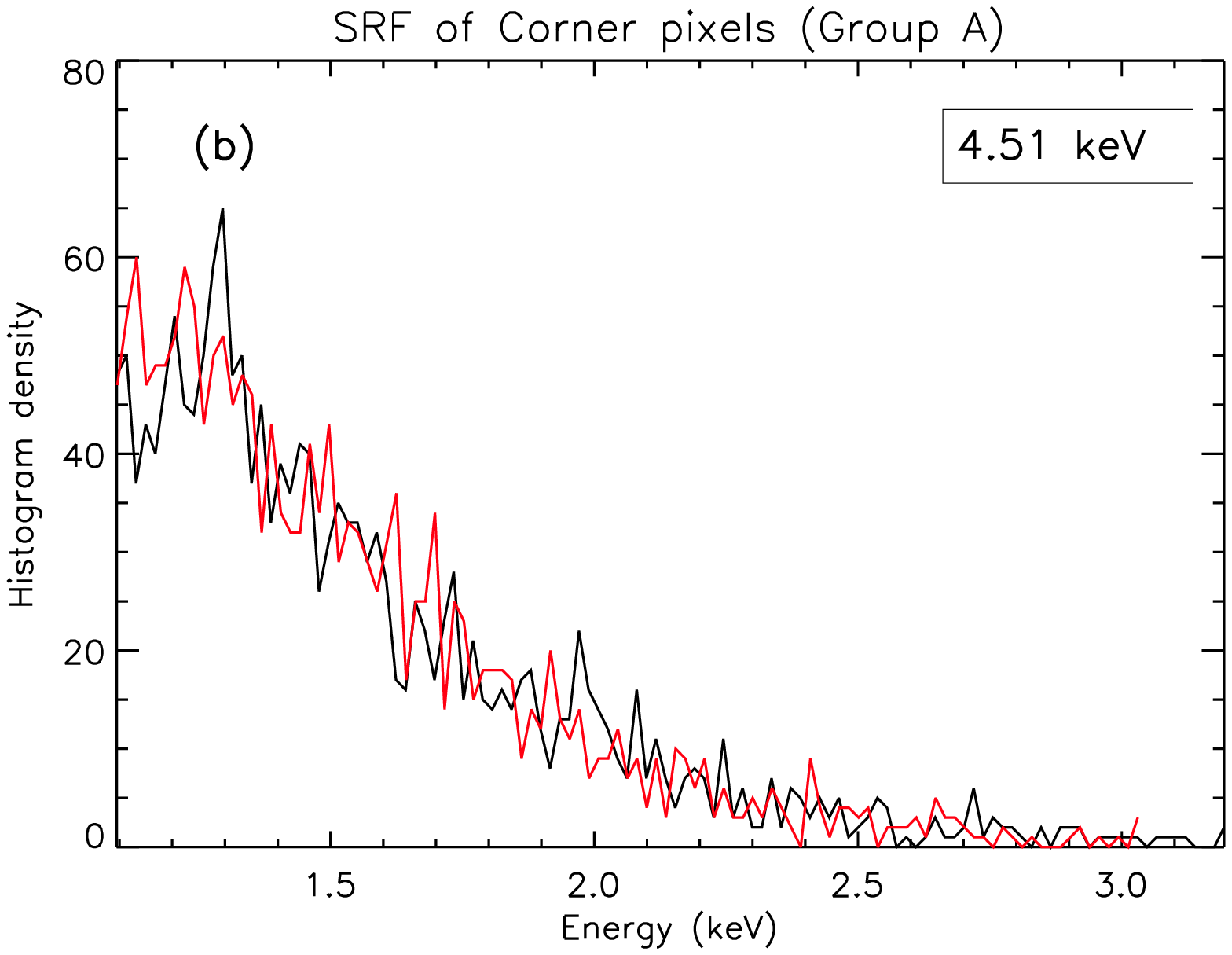}}
   \subfigure{\label{fig:ff_cornB}\includegraphics[height=4cm]{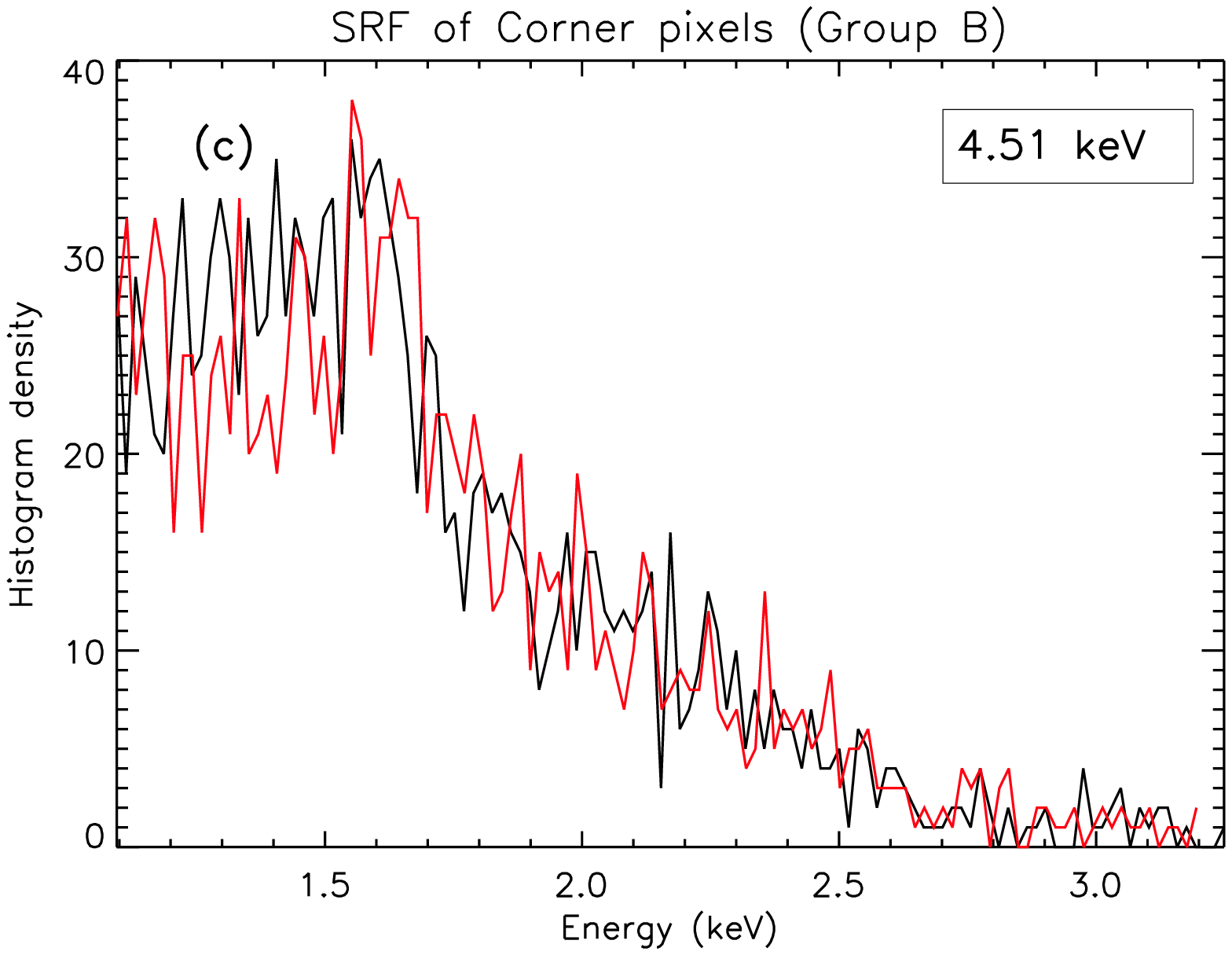}}
   \end{tabular}
   \end{center}
   \caption[example]
   { \label{fig:srffieldfreezone}
SRF of CCD-54 at 4.51 keV, simulated for interactions in field-free zone (a)
photons incident on center pixels (b) photons incident on group A corner
pixels (c) photons incident on group B corner pixels. Diagonal clocking of
wide spread split events is responsible for this SRF. The difference in SRF
between group A \& B is mainly due to diagonal readout of diffused charges
(i.e.,unlike group B,  group A pixels are along the diagonal readout). This clearly shows that field-free zone interactions contributes to the
low energy peak component in the total SRF.}
   \end{figure}

\subsection{Pixel dependency in Field zone}
\label{sec:field}  
Energy spectrum of photons interacting only inside
the depletion depth are discussed here. We also examine the SRF dependence
for corner and central pixels separately. Results of simulation using 4.51
keV X-rays are shown in Fig. \ref{fig:srffieldzone}. For non-corner pixels,
the SRF in this zone is characterized by a dominant photopeak (most of the
charge produced is swept-up by the anode) and a
low-energy tail. The low-energy tail contribution arises
from charge splitting due to interactions at pixel boundaries and
interactions at
greater depths. Due to charge loss
in the corner pixels (both group A \& B), its SRFs differ by the presence
of a dip
in the tail at low energies as shown in Fig. \ref{fig:f_corn}.

\subsection{Pixel dependency in Field-free zone}
\label{sec:fieldfree}  
Photon interaction in field-free zone is more complex.
The charge cloud diffuses into multiple pixels with some of the charges
lost due to recombination. The spectrum thus has a low energy distribution
without a photopeak. Energy response generated due to the
interactions in central pixels for 4.51 keV X-rays producing non-photopeak events are shown in
Fig. \ref{fig:ff_cent}. SRF of corner pixels from group A and B differ from each other as shown in Figs. \ref{fig:ff_cornA} \& \ref{fig:ff_cornB} . This difference in SRF
 could be because, the multi-pixel events due to diffusion in group A are in
 line with the diagonal readout which is not the case for group B. From
 Figs. \ref{fig:srffieldfreezone}(a, b, c) it is clear that the low energy peak component in the SRF is produced by the photon interactions in field-free zone.

\subsection{Energy dependency}
\label{sec:energy}  
\begin{figure}
   \begin{center}
   \begin{tabular}{c}
	   \subfigure{\label{fig:simulatedsrf}\includegraphics[height=6cm]{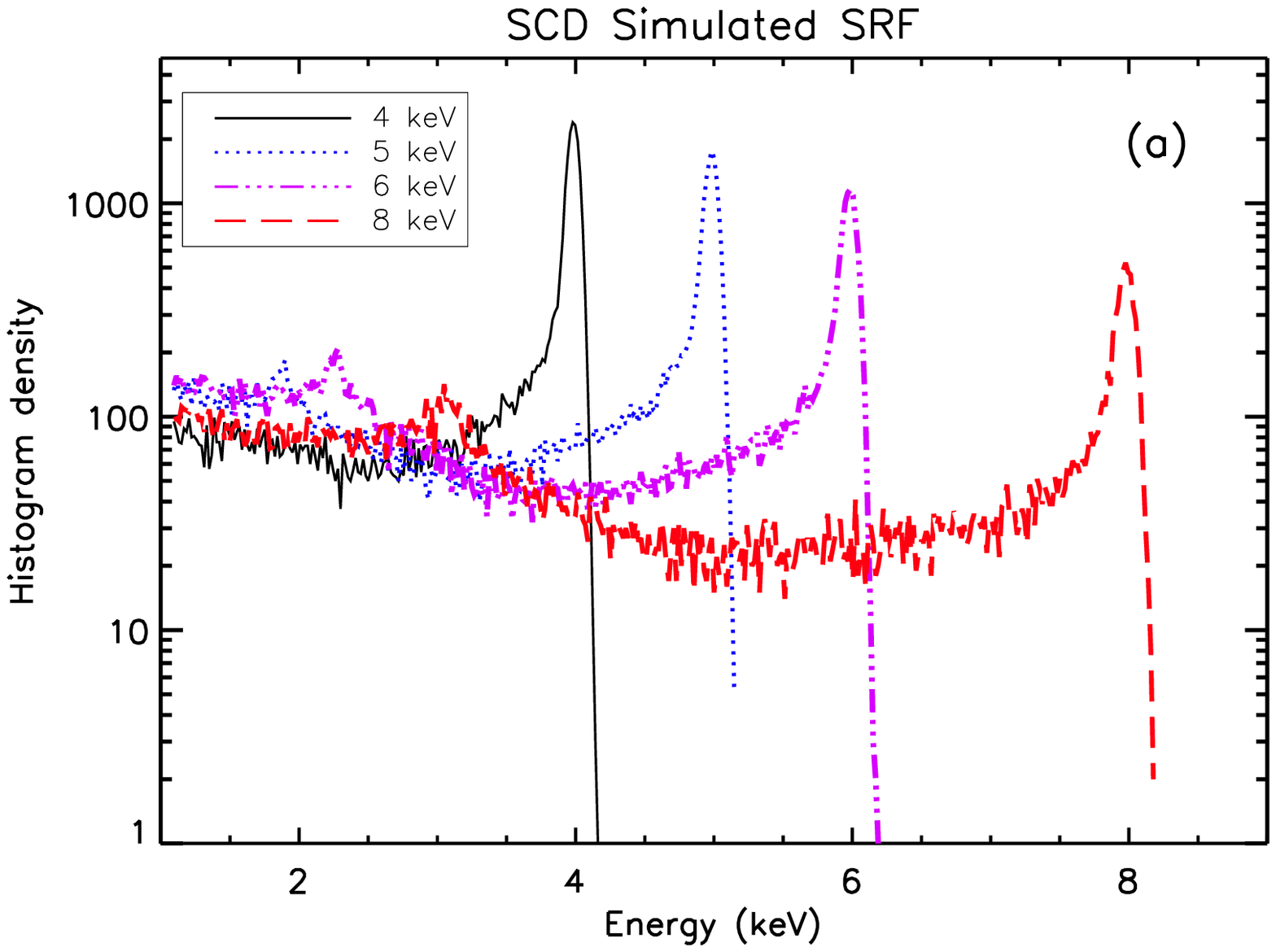}}
	   \subfigure{\label{fig:caldata_srf}\includegraphics[height=6cm]{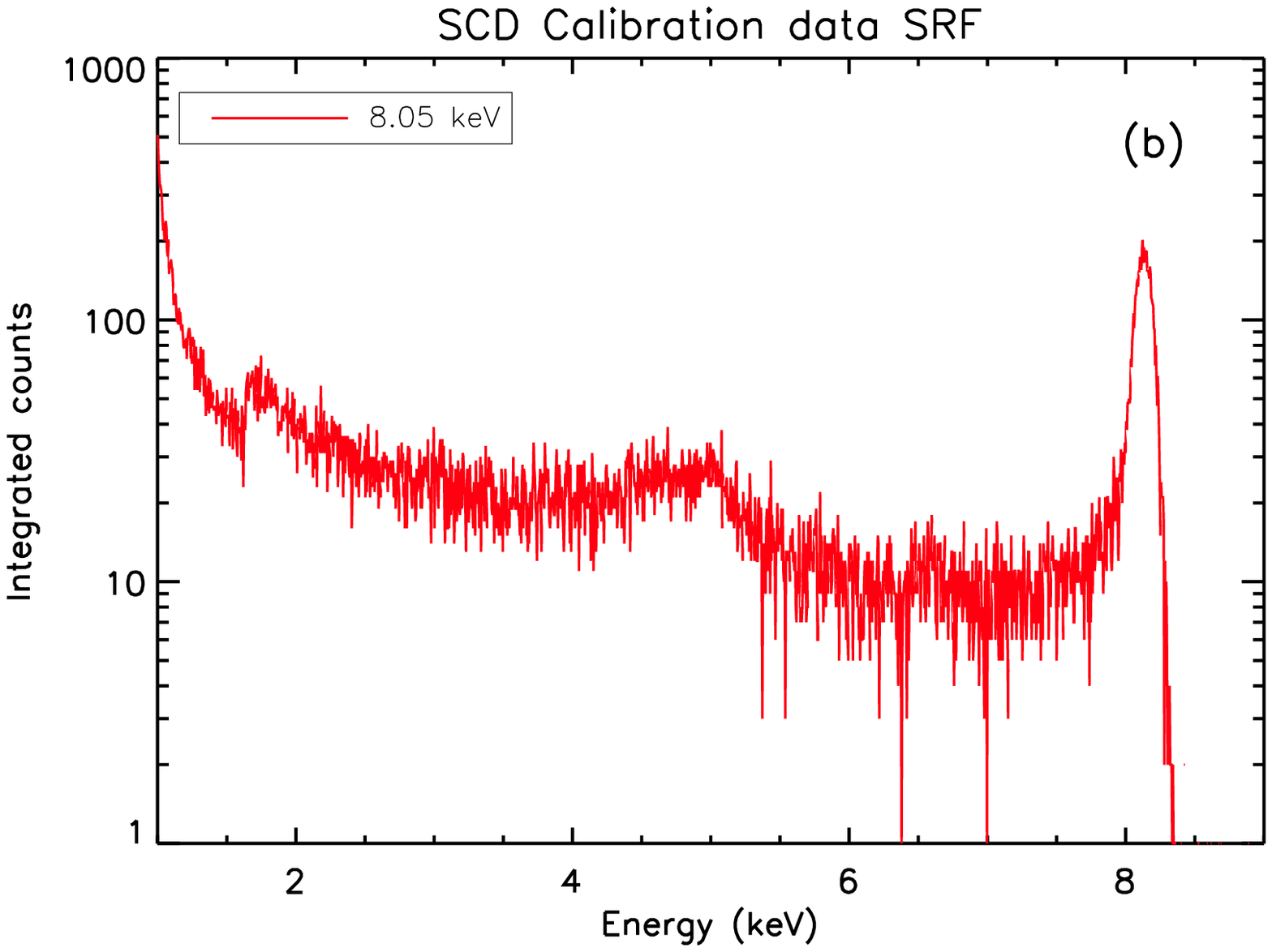}}
   \end{tabular}
   \end{center}
   \caption[example]
   { \label{fig:srfenergydep}
(a) SRF of CCD-54 simulated for different mono-energetic photons viz., 4 keV, 6 keV \& 8 keV
hitting a non-corner pixel in the detector (b) SRF at 8.05 keV of CCD-54
measured during C1XS calibration. Major SRF components viz., photopeak, low energy tail and low
energy peak are clearly seen in both the cases.}
   \end{figure}

\begin{figure}
   \begin{center}
   \begin{tabular}{c}
     \subfigure{\label{fig:ngauss_sim}\includegraphics[height=6cm]{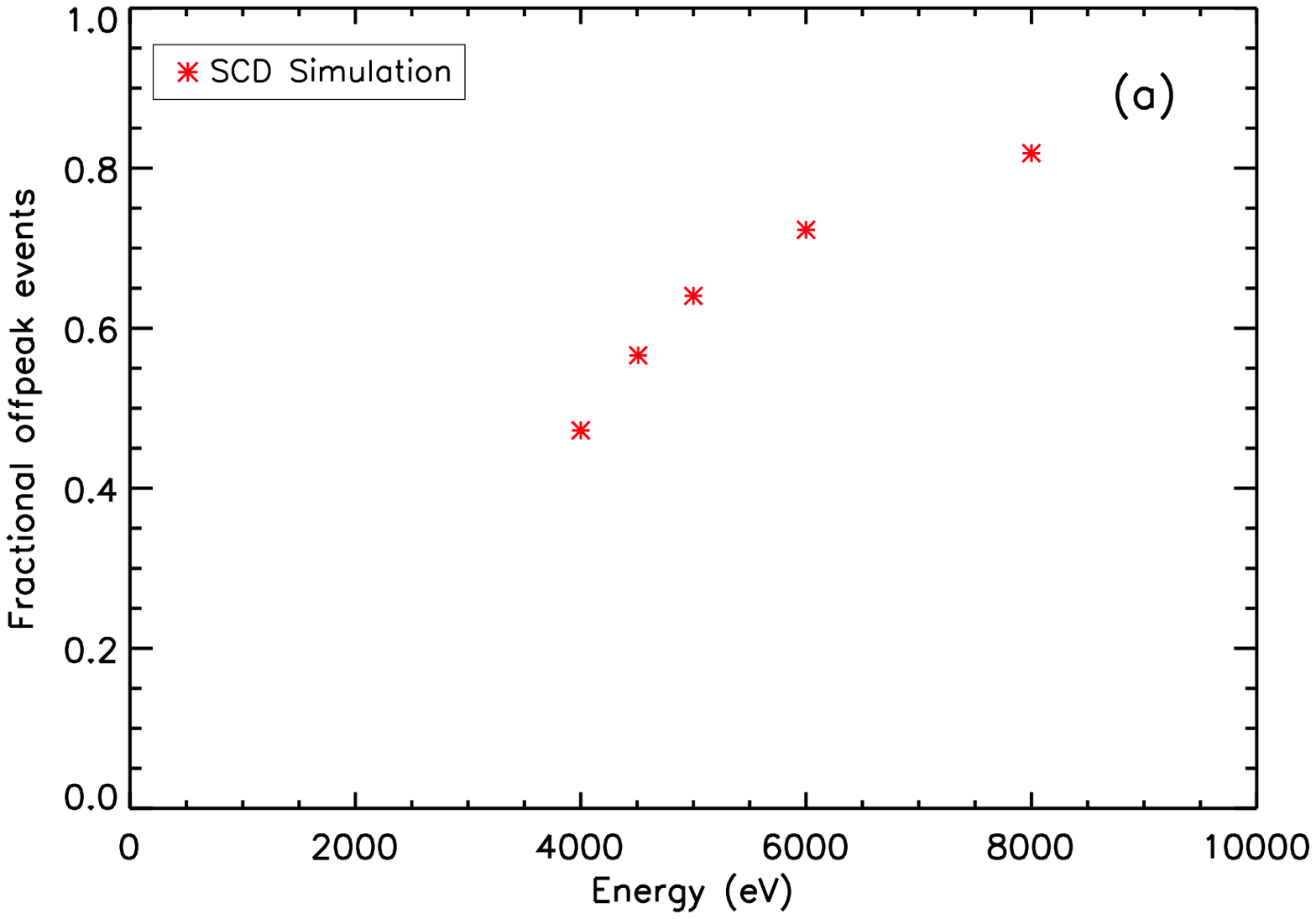}}
     \subfigure{\label{fig:ngauss_cal}\includegraphics[height=5.6cm]{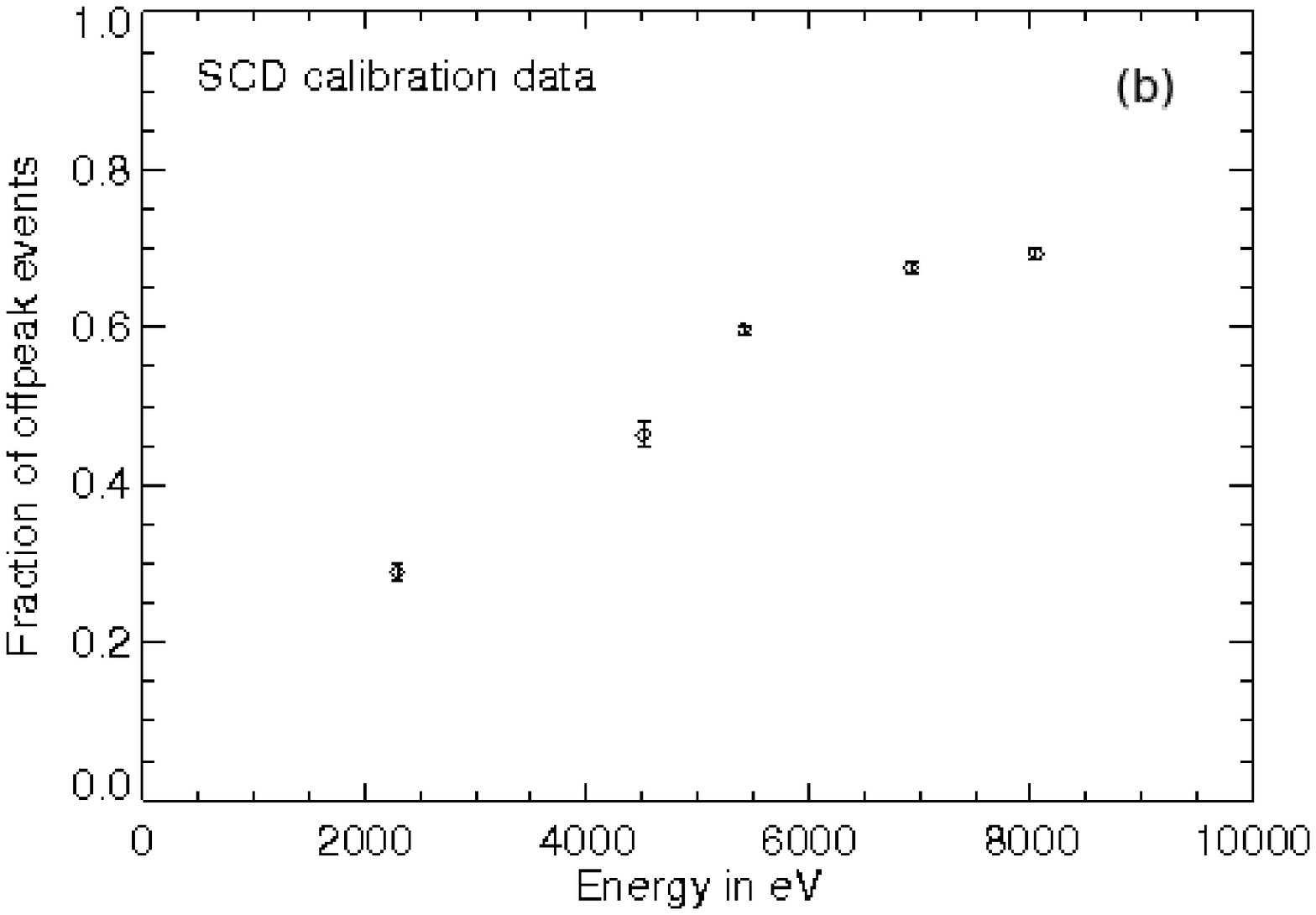}}
   \end{tabular}
   \end{center}
   \caption[example]
   { \label{fig:ngauss}
Variation of the off peak fractions in the SRF of CCD-54 as a function of energy is
plotted (a) simulation  (b) calibration data. The discrepancy between
the data and the model suggest to make refinements in the model.}
   \end{figure}

Here we present the energy dependence of SRF (i.e., 4 keV, 5 keV, 6 keV and 8 keV). Derived SRF containing the photopeak
and non-photopeak events (low energy tail and low energy peak) for X-rays incident on a center
pixel are shown in
Fig. \ref{fig:srfenergydep}. Features seen in the SRF are due to the
combined response from field
and field-free zone interactions.
High energy X-ray photons
penetrate deep inside the detector before interaction. Hence more
interactions occur very deep in the field zone and in often diffuse into
the field-free zone causing
more split events. The expected
trend of increase in photopeak and decrease in off-peak events with
decrease in X-ray energies is clearly seen. Fig. \ref{fig:caldata_srf} shows the spectrum of Cu K$_{\alpha}$ (8.05 keV)
measured with CCD-54 during C1XS ground calibration tests at the RESIK
facility, Rutherford Appleton Laboratory (RAL), UK . It is obvious from these plots that the
profile of the derived SRF from our current model reproduces fairly well
the calibration
data. Figure \ref{fig:ngauss_sim} and \ref{fig:ngauss_cal} establishes the  similarity in the variation of
off-peak component with energy between the derived SRF and that obtained
from laboratory calibration.\\


Calibration data shows that at 8.05 keV ${\approx}$ 75\% of total events are
non-photopeak events. While simulation result yields ${\approx}$ 82\% contribution for
the non-photopeak events in the total spectrum at 8keV.
The discrepancy could arise from non-inclusion of interactions in other zones
such as channel stop and also from assumption of ideal defect-free material
in the simulation model. Currently simulations
are performed only with small number of pixels and not with the exact
threshold logic employed in C1XS. Inclusion of interactions in dead layers, channel stop,
exact number of pixels and dimensions are all needed to establish a one-to-one
comparison between the two.



\section{Ongoing and Future work}
\label{sec:future}  
It is inferred from this work that
the proto-type model we have developed is promising and can give better
insights about charge transport in SCDs.
 A physical interpretation is given for the major
features in the SRF i.e., photopeak, and non-photopeak events (low energy
tail  \& low energy peak) (4 keV - 10 keV). The differences observed here clearly
suggest the need for further improvements in the current charge transport
model. Post model revisions, we plan to implement the event selection
logic used in C1XS to validate the model against data obtained during
extensive C1XS calibration. After validation, we will incorporate the
structure of CCD-236 with appropriate pixel dimensions in-order to generate
its SRF. Extensive tests are planned with CLASS to generate adequate
laboratory data on the dependencies of SRF using CCD-236.

\appendix    
\section{Equations for Field \& Field-free zone interactions } \label{sec:misc}

Equation used to compute the radius of charge cloud reaching the collection
node after drifting in the electric field is given by\cite{Hopkinson87}

	\begin{equation}
	\label{eq:fieldzoneradius}
	r_{f}  = 1 {\times} 10^4  \sqrt{\frac{4KT{\epsilon}}{e^2N_a}~ ln\left(\frac{d_d}{d_d -
	z_0}\right)}
	\end{equation}

where K is the Boltzmann constant, T is the temperature (in K), ${\epsilon}$ is the
electric permittivity of silicon, \textit{e} is the charge of an electron and $N_a$
is number density of acceptor impurities. For interaction in the epitaxial
field-free zone ($z_0 > d_d$), we assumed that no charge
enters into the bottom substrate layer i.e., charges are reflected back at
the boundary between field-free and substrate layer. We also assumed that
recombination is negligible as L ${\gg}$ $d_{ff}$. Equation to compute
the radius of charge cloud at the interface between field-free and
field zone is\cite{Pavlov99} :

\begin{equation}
	\label{eq:fieldfreezoneradius}
	r_{ff}  = \sqrt{2d_{ff}L\left [ tanh\left(\frac{d_{ff}}{L}\right) - \left(1-\frac{z_0
	-d_d}{d_{ff}}\right) tanh\left(\frac{d_{ff}-z_0+d_d}{L}\right)\right ]}
	\end{equation}

where  $d_d$ is the depletion depth,
$z_0$ is the interaction depth in the detector, $d_{ff}$ is the thickness
of epitaxial field free zone, L is the diffusion length.

Final radius of the charge cloud is given by :

\begin{equation}
	\label{eq:finalradius}
	r  = \sqrt{r_i^2+r_d^2+r_{ff}^2}
	\end{equation}

Assuming Gaussian charge density profile for the cloud in both field and
field-free zone cases, the equation to derive the amount of charges
collected in each pixel (i,j) is given by Pavlov and Nousek\cite{Pavlov99}:
\begin{equation}
	\label{eq:chargecollected}
	Q_{ij}(x0,y0) = \frac{Q_0}{4}
	\left[erf \left (\frac{a_{i+1}-x_0}{r}\right ) -
	      erf \left (\frac{a_{i}-x_0}{r} \right) \right ]
	\left[erf \left (\frac{b_{j+1}-y_0}{r}\right ) -
	      erf \left (\frac{b_{j}-y_0}{r} \right) \right ]
	\end{equation}
	where $Q_{0}$ is the initial charge ($E_f$/${\omega}$), a and b are pixel dimensions, $a_i$ = -$\frac{a}{2}$+ia, $b_j$ =
	-$\frac{b}{2}$+jb and r is given by Eq. \ref{eq:finalradius}.
	$x_0$, $y_0$ are photon interaction coordinates in a pixel where photon
	is absorbed with its origin at the pixel center (i.e., -a/2 $<
	x_0 <$ a/2, -b/2 $<y_0 <$ b/2).

\end{document}